\documentstyle[aps,multicol,psfig]{revtex}
\newcommand{\bd}[1]{ \mbox{\boldmath $#1$}  }

\title{Cross sections for Coulomb and nuclear breakup of three-body 
halo nuclei}

\author{E. Garrido}
\address{Instituto de Estructura de la Materia, CSIC, Serrano 123, E-28006
Madrid, Spain}
\author{D.V.~Fedorov and A.S.~Jensen}
\address{Institute of Physics and Astronomy,
Aarhus University, DK-8000 Aarhus C, Denmark}
\date{\today}

\begin{document}
\draft

\maketitle

\begin{abstract} 
All possible dissociation cross sections for the loosely bound
three-body halo nuclei $^6$He (n+n+$\alpha$) and $^{11}$Li
(n+n+$^{9}$Li) are computed as functions of target and beam
energy. Both Coulomb and nuclear interactions are included in the same
theoretical framework. The measurements agree with the
calculations for energies above 100 Mev/nucleon. The largest cross 
sections correspond to final states
with zero or three particles for heavy and with two neutrons for light
targets.
\vspace{1pc}
\end{abstract}


\widetext
\tighten
\begin{multicols}{2}

\paragraph*{Introduction.}
 
Halo nuclei are spatially extended bound systems
\cite{rii94,han95,tan96,jon98}. Their existence is revealed by large
total interaction cross sections \cite{tan96,ber93}. The often small
one or two-nucleon separation energies led to a successful description
in terms of two or three-body structures \cite{zhu93}. This in turn
emphasizes the importance of reactions breaking these halo
structures into their constituent particles. We shall here concentrate
on two-neutron halos where the three-body problem is then
inherent. Absolute values for dissociation cross sections of these
systems are available in a number of cases
\cite{tan88,kob89,bla93,hum95,zin97,ale98,aum99}. They
may be divided into processes like one or two neutron knockout arising
from Coulomb or nuclear dissociation. Also core destruction processes
were studied \cite{nil95} and the total interaction cross section
obtained \cite{tan96}.

Theoretical investigations of  two-neutron dissociation cross
sections are available on light \cite{suz94,ber98,gar99} as well as
heavy \cite{ber93,oga92,cob98} targets, but core destruction processes
are usually not computed at all. Total interaction cross sections
for light targets are also calculated for three-body halos
\cite{suz94,alk96}. However, systematic studies of all these processes
within one consistent model are not available. Such results even for
relative quantities are rather scarce \cite{gar99}.  Reliable nuclear
model estimates of absolute values are in general very difficult. In
particular the necessary simultaneous treatment of the Coulomb and
nuclear interactions has not been available for these three-body halo
reactions.

The lack of systematic experimental and theoretical information about
the many different absolute dissociation cross sections is perhaps
surprising, but in any case unfortunate, since characteristic features
of the reaction mechanism probably can be uncovered by such
investigations.  The purpose of this letter is then to study all
possible three-body dissociation cross sections of two-neutron halo
nuclei as functions of target and beam energy. The predictions employ
a consistent three-body model and an appropriate reaction framework
including simultaneous treatment of Coulomb and nuclear interactions.

\paragraph*{Theoretical formulation.}

We shall here only give a brief sketch of the model leaving the
details for a more comprehensive publication. We want to describe
collisions of a target and a weakly bound two-neutron halo nucleus
and classify the cross sections according to the particles left in 
the final state, i.e. $\sigma _{nnc}$, $\sigma_{nc}$, $\sigma _{nn}$, 
$\sigma _{c}$, $\sigma _{n}$, $\sigma _{0}$,
where $c$ and $n$ denote core and neutron, respectively and $0$
indicates that all particles are absorbed.

The collisions can approximately be described by
three independent collisions of one particle (participant) at a time
and undisturbed motion of the remaining two projectiles particles
(spectators) \cite{gar99}. The interconnected main assumptions are
that the halo is weakly bound, spatially extended and the intrinsic
halo motion is slow compared to the relative projectile-target motion.

In this simple picture the spectators are always found in the final
state while the participant can be either absorbed or scattered by
the target. However the finite extension of the projectile constituents
and the target is destroying this simple picture, since simultaneous
collisions of more than one projectile constituent with the target are
possible.

The simplest description of the constituent-target interaction can be
made through the black sphere model. This model assumes that a
particle is absorbed when passing inside a cylinder with the axis
along the beam direction and left untouched otherwise. The radius of
this cylinder is approximately equal to the sum of the target and
constituent radii.  Different contributions to the cross section have
to be considered, $i)$ three contributions where only one of the
constituents is inside the cylinder (one participant and two
spectators). In this case the interaction between the participant and
the target is described by the phenomenological optical model, and
only the part of the three-body wave function where the two spectators
are far enough from the participant is included \cite{gar99}. $ii)$
Three contributions where two constituents are inside the
cylinders. In this case, the interaction between one of them
(participant) and the target is described by the optical model, while
for the second constituent-target interaction we use the black sphere
model. To implement this we include only the part of the wave function
where the second constituent is close to the participant and the third
one is far enough away. $iii)$ One contribution where the three
constituents are inside the cylinders. Again the interaction of one of
the constituents (participant) with the target is described by the
optical model, and for the other two we use the black sphere model
(only the part of the wave function where the three particles are
close enough is included).  It is then clear that the distances
between the participant and the other two constituents appear as the
decisive quantities determining if more than one particle interacts
simultaneously with the target. These distances depend on the sizes of
the constituents and the target and they are determined as in \cite{gar99}.

In the contributions to $ii)$ and $iii)$ the participant has to be
chosen, and the participant-target interaction is described by the 
optical model. Since the core-target interaction
contains nuclear and Coulomb forces a more careful treatment is
required in this case. Therefore the contributions arising from
simultaneous interactions of the core and one or two neutrons with the
target are computed with the core as participant and the core-target
interaction described by the optical model.

The division into different contributions implicitly assumes that for
large impact parameters the constituent and the target are not
interacting. Obviously this is not true for the long-range Coulomb
interaction. To solve this problem, when the core is the participant, we 
include into the contributions
$i)$, $ii)$ and $iii)$ only the large momentum transfer part (low impact
parameter) of the Coulomb interaction. The low momentum transfer
contribution (large impact parameter) is considered as a process where
the core is elastically scattered by the target (via Coulomb
interaction) and the two neutrons survive untouched in the final
state. The value of the momentum transfer dividing into low and large
impact parameters is given by \cite{ber88} $q_g = Z_0Z_ie^2 (\gamma
+1)/(c \gamma \beta (R_0+R_i+ \pi a /2))$. Here $R_0$ and $R_i$ are
charge root mean square radii of the target and the core participant 
($^9$Li or $^4$He) and $a$ is half the distance of
closest core-target approach, $eZ_0$ and $eZ_i$ are the charges of the
target and participant, $\beta = v/c$ and $\gamma =
1/\sqrt{1-\beta^2}$.
The momentum cutoff parameter $q_g$ separates between impact parameters
smaller and larger than the sum of participant and target radii. In the
present context this means distinction between absorption (destruction) 
and survival of the participant.

The differential cross sections arising from participant $i$ with mass
$m_i$ and charge $eZ_i$ has contributions from elastic scattering
$\sigma _{el}^{(0i)}$ (diffraction) and absorption $\sigma _{abs}^{(0
i)}$ (stripping) on the target. For a spinless target of mass $m_0$
and charge $eZ_0$ we get in the rest system of the halo
\begin{eqnarray} \label{eq6}
 \frac{d^6\sigma _{abs}^{(i)}(\bd{p}_{0i,jk}^{\prime },\bd{p}_{jk}^{\prime })}
{ d\bd{p}_{0i,jk}^{\prime } d\bd{p}_{jk}^{\prime } }
=   \sigma _{abs}^{(0i)}(p_{0i}) \; 
 |M_s(\bd{p}_{i,jk}, \bd{p}_{jk}^{\prime })|^2 \; , 
\end{eqnarray}
where $M_s$ is the normalized overlap matrix element between initial
and final state spectator wave functions, $\bd{p}_{0i,jk}^{\prime }$
is the relative momentum in the final state between center of mass of
target-participant and the spectators $j$ and $k$, while
$\bd{p}_{jk}^{\prime }$, $\bd{p}_{0i}$ and $\bd{p}_{i,jk}$
correspondingly are relative momenta between particles $j$ and $k$,
$0$ and $i$, $i$ and center of mass of $j$ and $k$. Primes denote
final states. Momentum conservation in the rest frame of the
projectile gives the relation $\bd{p}_{0i,jk}^{\prime } =
\bd{p}_{i,jk} + \bd{p}_{0} (m_j+m_k) / (m_0+m_i+m_j+m_k)$, where
$\bd{p}_{0}$ is the momentum of the target.

The differential elastic cross section is
\begin{eqnarray}
&&  \frac{d^9\sigma _{el}^{(i)}(\bd{p}_{0i,jk}^{\prime },
\bd{p}_{jk}^{\prime },\bd{p}_{0i}^{\prime })}
{ d\bd{p}_{0i,jk}^{\prime } d\bd{p}_{jk}^{\prime } d\bd{p}_{0i}^{\prime } }
  =   \frac{d^3\sigma _{el}^{(0i)}(\bd{p}_{0i}
  \rightarrow  \bd{p}_{0i}^{\prime})} {d\bd{p}_{0i}^{\prime }} 
\nonumber \\ && \; \; 
\times  \left( 1 - |\langle \Psi| \exp(i \delta {\bd q}
 \cdot  \bd{r}_{i,jk}) |\Psi \rangle|^2 \right)
 |M_s(\bd{p}_{i,jk}, \bd{p}_{jk}^{\prime })|^2 \; ,  \label{eq5}
\end{eqnarray}
where $\Psi$ is the initial three-body halo state. We have now a
9-dimensional differential cross section, since the participant
explicitly is included in the final state.  The momentum $\delta {\bd
q} = (\bd{p}_{i,jk}^{\prime } - \bd{p}_{i,jk})
(m_j+m_k)/(m_i+m_j+m_k)$ is the transfer into the
participant-spectators relative motion described by the coordinate
$\bd{r}_{i,jk}$. The second factor in eq.(\ref{eq5}) then expresses 
the probability for
the halo {\it not} ending up in its ground state. In this way we
remove elastic scattering of the halo as a whole.

When the participant is charged the Coulomb interaction produces a
logarithmic divergence in the total cross section Eq.(\ref{eq5}). The
corresponding adiabatic motion related to virtual excitations at large
impact parameters should be removed from the dissociation cross
sections \cite{ber88}. We therefore exclude contributions from
momentum transfer smaller than the adiabatic cutoff $q_a = \hbar
Z_0Z_i e^2 /(\pi \hbar c) \; B_{ps} /(\hbar c) \; (\gamma +1)
\gamma^{-2} \beta^{-2}$, where $B_{ps}$ is the binding energy between
participant and the system consisting of the spectators, i.e., 
$B_{ps} = B - B_{2s}$, where $B$ is
the three-body binding energy and $B_{2s}$ is the two--body binding energy
of the two spectators. Note that in a Borromean nucleus $B_{2s}$ is
negative. 

The energy transferred from target to participant, $ \delta E \equiv
\sqrt{{\bf p}_0^2 + m_0^2} - \sqrt{{\bf p_0^\prime}^2 + m_0^2}$, must
be larger than $B$. When ${\bf p_0}$ and ${\bf
q} \equiv {\bf p_0} - {\bf p_0}^\prime$ are parallel $\delta E$ is
maximized. For this geometry we find for small $B$ compared to the
target rest mass that $\delta E = B$ implies that $q c \equiv q_L c
\approx B \sqrt{1+m_0^2c^2/p_0^2}$ which reduces to $B/v$ in the
non-relativistic limit. Thus $q$ must be larger than $q_L$ to produce
dissociation, but on the other hand dissociation is not the necessary
outcome for all $q > q_L$.  In the computations we exclude
contributions from momentum transfer $q$ smaller than the largest of
$q_L$ and $q_a$.

\paragraph*{Cross sections.}

The model is now completely defined with both Coulomb and nuclear
interactions included for weakly bound three-body halo reactions. We
shall study breakup reactions of $^6$He and $^{11}$Li on C, Cu, and Pb
targets. The parameters corresponding to the $^6$He and $^{11}$Li wave
functions are obtained from \cite{gar99,cob98}. The optical model
parameters are from \cite{coo93} for neutrons, from \cite{nol87} for
$\alpha$-particles and for $^9$Li also from \cite{nol87} but using
range and diffuseness parameters from \cite{zah96}.  We furthermore
drastically reduce the energy dependence of the real part of the
potential in \cite{nol87}, i.e. $a_2 = -0.014$, to allow for the
required huge beam energy variation. The measured core-target
interaction cross sections are reproduced within error bars
\cite{bla93}.  The binding energy $B_{ps}$ between the $^{9}$Li and
$^4$He cores and the two neutrons must be introduced for the adiabatic
cutoff. We use the scaling relation in \cite{joh90} to obtain
$B_{ps}/B \approx 3$ for $^6$He and 1.4 for $^{11}$Li.

The corresponding contributions to the dissociation cross sections are
obtained as indicated in $i)$, $ii)$ and $iii)$ by integration of
Eqs.(\ref{eq6}) and (\ref{eq5}). The cross sections are then
classified according to the particles left in the final state,
i.e. $\sigma _{nnc}$, $\sigma_{nc}$, $\sigma _{nn}$, $\sigma _{c}$,
$\sigma _{n}$, and $\sigma _{0}$. Specific interesting cross section
combinations are those of two-neutron removal $\sigma _{-2n} \equiv
\sigma _{nnc} + \sigma _{nc} + \sigma _{c}$, core destruction $\sigma
_{-c} \equiv \sigma _{nn} + \sigma _{n} + \sigma _{0}$ and the sum of
these, the total interaction cross section $\sigma _{I} \equiv \sigma
_{-2n} + \sigma _{-c}$.

\begin{figure}[t]
\centerline{\psfig{figure=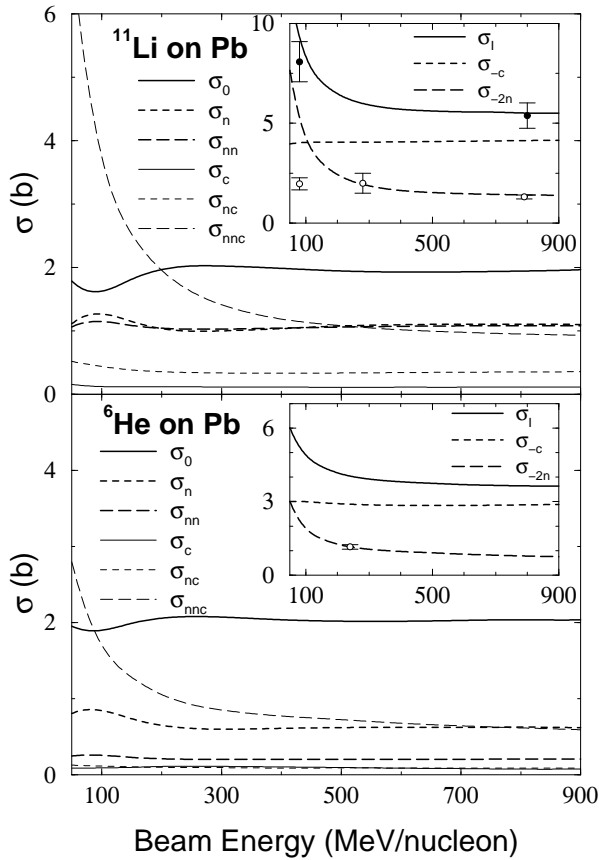,width=7.8cm,%
bbllx=2.4cm,bblly=2.4cm,bburx=18.8cm,bbury=26.4cm,angle=0}}
\vspace{0.5cm}
\caption[]{Dissociation cross sections as function of beam energy for 
fragmentation of $^{11}$Li (upper part) and $^6$He (lower part) on a 
Pb-target. The labels indicate the halo particles in the final state.
The inset shows  $\sigma_{-2n} =
\sigma_{c} + \sigma_{nc} + \sigma_{nnc}$, $\sigma_{-c} = \sigma_{0} +
\sigma_{n} + \sigma_{nn}$, $\sigma_{I} = \sigma_{-2n} + \sigma_{-c}$. 
The experimental data are from \protect\cite{kob89,bla93,zin97,aum99}.}
\label{fig1}
\end{figure}

In fig. \ref{fig1} we show the results for a lead target and $^{11}$Li
(upper part) and $^6$He (lower part) projectiles.  At low beam
energies $\sigma_{nnc}$ is the dominant cross section, especially in
the $^{11}$Li case. This is due to the large Coulomb interaction, that
highly increases the large impact parameter contribution (where the
two neutrons both survive 
\begin{figure}
\centerline{\psfig{figure=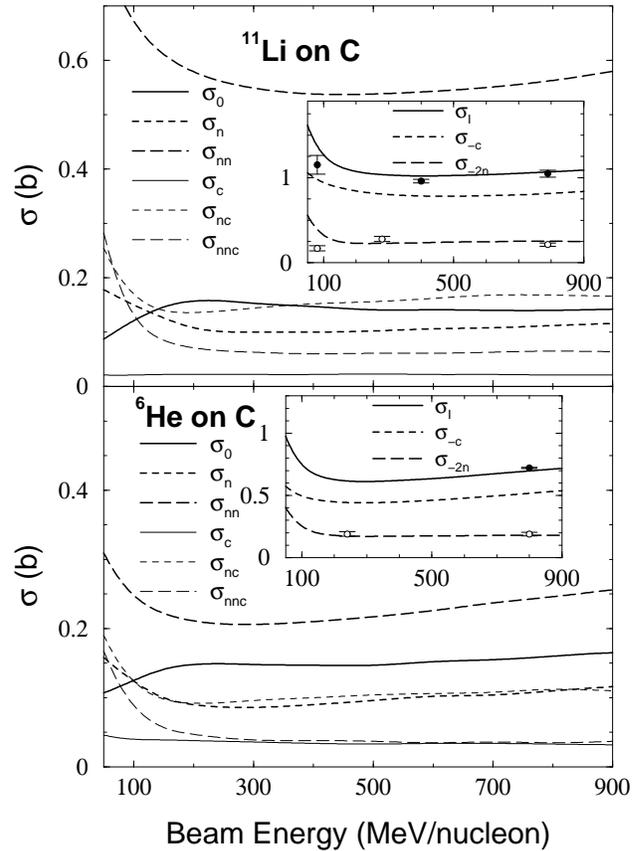,width=7.8cm,%
bbllx=2.4cm,bblly=2.4cm,bburx=18.8cm,bbury=26.4cm,angle=0}}
\vspace{0.5cm}
\caption[]{The same as Fig.\ref{fig1} for a C-target. The 
experimental data are from \protect\cite{tan88,bla93,zin97,aum99,suz94}.}
\label{fig2}
\end{figure}
\noindent
in the final state). The $\sigma_{nnc}$ cross section rapidly decreases 
with the beam energy, and for energies
larger than 200 MeV/nucleon in the $^{11}$Li case and 100 MeV/nucleon
in the $^6$He case $\sigma_0$ dominates (no particles in the final
state).  The reason for this is the large radius of the Pb-target,
resulting in a very high probability for finding all the three
projectile constituents inside the absorption cylinders.  This
probability is higher for $^6$He projectile than for $^{11}$Li,
because from the three-body wave function we get $\langle r_{cn}^2
\rangle^{1/2}=4.2$ fm in the first case and 5.9 fm in the second
($r_{cn}=$ neutron-core distance). As a consequence, when the core is
the participant, 85\% of the $^6$He wave function corresponds to all
three constituents inside the cylinders and only 65\% for the
$^{11}$Li projectile. This is reflected in the figure by the fact that
$\sigma_0$ takes very similar values in both cases, while as a general
rule the cross sections for $^{11}$Li should be larger than the ones
for $^6$He. The core destruction cross section $\sigma_{-c}=\sigma_{0}
+ \sigma_{n} + \sigma_{nn}$ is shown in the insets by the short-dashed
line. Its behavior is determined by the core--Pb absorption cross
section, and changes very little with the beam energy. The two-neutron
removal cross section $\sigma _{-2n} = \sigma _{nnc} + \sigma _{nc} +
\sigma _{c}$ is shown as the long-dashed lines in the insets, and it
is given by all the processes where the core survives and contains
therefore the contribution from the Coulomb interaction. Therefore
this cross section decreases with 
beam energy. Finally the solid lines in the insets show the
\begin{figure}
\centerline{\psfig{figure=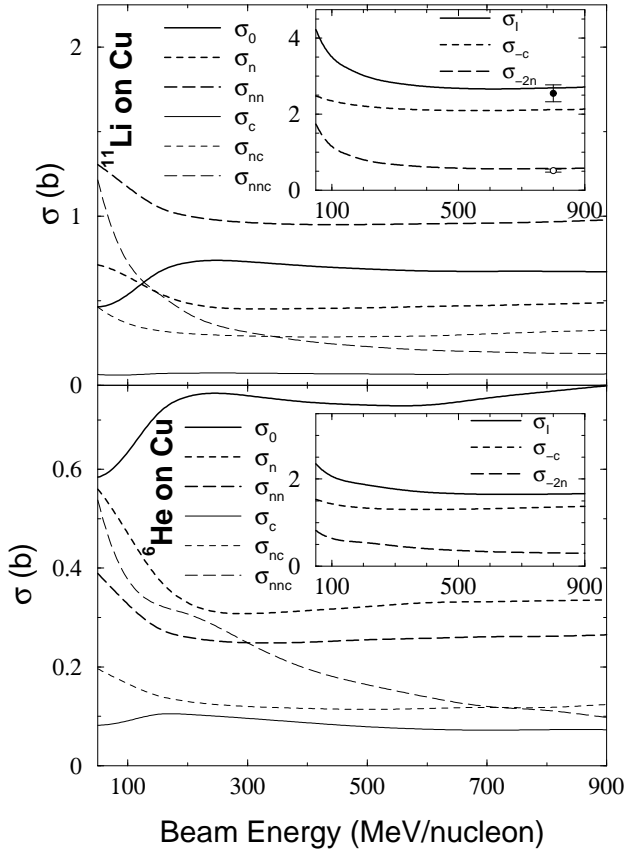,width=7.8cm,%
bbllx=2.4cm,bblly=2.4cm,bburx=18.8cm,bbury=26.4cm,angle=0}}
\vspace{0.5cm}
\caption[]{The same as Fig.\ref{fig1} for a Cu-target. The 
experimental data are from \protect\cite{kob89}.}
\label{fig3}
\end{figure}
\noindent
interaction cross section $\sigma _{I} = \sigma
_{-2n} + \sigma_{-c}$. The agreement with the experimental data is
remarkably good.

In fig. \ref{fig2} we show the results for a light target (carbon)
where the cross sections are up to a factor of 10 smaller than for a
lead target. The main difference is that the effect produced by the
Coulomb interaction is now small, and for example $\sigma_{-2n}$ is 
determined by Coulomb for Pb and by nuclear interactions for C
targets. The dominating cross section is now $\sigma_{nn}$ for C
instead of $\sigma_{nnc}$ or $\sigma_{0}$ for Pb targets. The reason
is that the small radius of the target diminishes the probability of
simultaneous interaction of more than one constituent with the target,
and absorption of the core together with two truly undisturbed
spectator neutrons becomes the most likely process. The variation with
beam energy is in general rather similar to that of Pb, in particular
revealed by $\sigma_{-c}$, $\sigma_{-2n}$ and $\sigma_I$ shown in the
insets. Also in this case we obtain good agreement with the
experimental data.

In fig. \ref{fig3} we show the results for a copper target, i.e. an
intermediate mass with comparable Coulomb and nuclear contributions.
The cross sections are between those obtained for carbon and lead with
similar energy dependences.  However, $^{11}$Li has a larger radius
than $^6$He.  Thus the dominating cross section for $^{11}$Li is
$\sigma_{nn}$ as for carbon, while $\sigma_0$ dominates for $^6$He as
for a lead target.

Only a few of the computed cross sections are experimentally available.
In \cite{aum99} experimental data for $\sigma_{nnc}$, $\sigma_{nc}$ and
$\sigma_c$ are given for $^6$He on C and Pb at 240 MeV/nucleon.
In \cite{zin97} the experimental data for the same cross sections are
given for $^{11}$Li fragmentation on C and Pb at 280 MeV/nucleon.
Also in \cite{ber98} $\sigma_{nnc}$, $\sigma_{nc}$ and $\sigma_c$
are calculated for $^{11}$Li and  $^6$He on C by use of the eikonal
theory. In table \ref{tab1} we compare the results of our calculation
with the ones given in \cite{ber98} and the experimental data in
\cite{zin97,aum99}. For carbon target the agreement with the experimental
data and previous calculations is good. The only significant 
deviation is 
in $\sigma_{nc}$ for the case of lead target, where our
calculation gives a cross section at least a factor of two smaller 
than the experimental value.

\end{multicols}

\begin{table}[b]
\begin{tabular}{ccccc|ccc}
 & & \multicolumn{3}{c}{$^6$He (240 MeV/nucleon)} & 
              \multicolumn{3}{c}{$^{11}$Li (280 MeV/nucleon)} \\ 
Target & & $\sigma_{nnc}$ & $\sigma_{nc}$ & $\sigma_{c}$ &
           $\sigma_{nnc}$ & $\sigma_{nc}$ & $\sigma_{c}$  \\ \hline
 & Exp.\protect\cite{zin97,aum99} &
  $30\pm 5$ & $127\pm14$ & $33\pm23$ & $60\pm20$ & $170\pm20$
                                    & $50\pm10$  \\
C  & This work & 43 & 93 & 37 & 64 & 142 & 22 \\
 &Ref.\protect\cite{ber98}& 32 & 136 & 17 & 6--10 & 121--162 & 27--40 \\ \hline
Pb & Exp.\protect\cite{zin97,aum99} &
  $650\pm110$ &  $320\pm90$ & $180\pm100$ & $1000\pm350$ & $1000\pm350$
                                    & $\leq 70 $  \\
 & This work & 940 & 96 & 113 & 1476 & 338 & 116 
\end{tabular}
\caption{Computed values of $\sigma_{nnc}$, $\sigma_{nc}$ and $\sigma_c$
for $^6$He on C and Pb at 240 MeV/nucleon and for $^{11}$Li on C and
Pb at 280 MeV/nucleon. For comparison we give the results reported 
in \protect\cite{ber98}. Experimental data from
\protect\cite{zin97,aum99}.}
\label{tab1}
\end{table}

\begin{multicols}{2}

The core destruction processes entering in $\sigma_{-c}$ all must take
place at small impact parameters. The Coulomb contributions
are therefore relatively unimportant in contrast to the large impact 
parameter processes where purely Coulomb dissociation reactions
take place. Coulomb dissociation cross sections have been previously
estimated in models where the three--body system is treated as
an effective two--body (dineutron plus core). For $^{11}$Li on
lead at 800 MeV/nucleon we get 765 mb, while in \cite{joh90}
they obtain 960 mb, and in \cite{suz90} they report values ranging
from 610 to 660 mb. For $^{11}$Li on Cu at 800 MeV/nucleon we obtain
a Coulomb dissociation cross section of 81 mb, while in \cite{suz90}
they give values from 86 to 92 mb. Calculations where the final
continuum three--body wave function is considered are also available
\cite{cob98}. For $^{11}$Li on Pb at 180, 280, and 800 MeV/nucleon
they get 2128 mb, 1429 mb and 971 mb, respectively, while our 
calculations for the same energies give 2050 mb, 1401 mb and
765 mb. These rather few previous computations where a comparison
is possible are in general in rough agreement with our systematic results.

\paragraph*{Conclusion.}

We have computed all possible three-body dissociation cross sections
of two-neutron halos as function of beam energy and target. Coulomb
and nuclear interactions are treated within the same framework. The
available experimental information, mostly about total interaction and
two-neutron separation cross sections, compares rather well with the
calculated results, especially at high energies.  At low
energies the three--body continuum final state might turn out to be 
more appropriate. The further division into a specific number of
particles in the final states carries detailed information about the
reaction mechanism. We predict the absolute sizes of all these cross
sections to encourage new measurements.

\paragraph*{Acknowledgement.} We thank K. Riisager for continuous 
discussions and suggestions.

\end{multicols}

\end{document}